\documentclass[12pt, a4paper]{article}
\usepackage{amsmath}
\usepackage{amsfonts}
\usepackage{amssymb}
\usepackage{graphicx,epsfig}
\usepackage{cite}
\usepackage{upgreek}
\usepackage{textcomp}
\begin{document}
\title{Charge Motion along Polynucleotide Chains in a Constant Electric Field Depends on the Charge Coupling Constant with Chain Displacements}

\author{A.N. Korshunova, V.D. Lakhno \\

\small{Institute of Mathematical Problems of Biology RAS --}\\
\small{the Branch of Keldysh Institute of Applied Mathematics RAS}\\

\footnotesize{E-mail: alya@impb.ru (A.N. Korshunova); lak@impb.ru (V.D. Lakhno)}}
\date{}

\maketitle

\begin{abstract}
	Various regimes of a charge motion along a chain in a constant electric field are investigated. This motion is simulated on the basis of the Holstein model. Earlier studies demonstrate a possibility of a uniform motion of a charge in a constant electric field over very long distances. For small values of the electric field intensity a Holstein polaron can move at a constant velocity. As the electric field intensity increases, a charge motion acquires oscillatorily character, performing Bloch oscillations. Since the charge motion depends on the whole set of the system parameters the character of the motion depends not only on the value of the electric field intensity. Therefore, the electric field intensity for which the uniform motion takes place differs for chains with different parameters. The character of the charge motion and distribution is considered in chains with different values of the constant of coupling between the charge and the displacements of the chain.
	It is shown that the values of the electric field intensity for which the regime of a charge motion changes are different in chains with different values of the coupling constant. It is also demonstrated that for one and the same value of the electric field intensity, in chains with different values of the coupling constant either a uniform motion or an oscillatory motion, or a stationary polaron can be observed.
\end{abstract}

\textbf{Keywords:} nanobioelectronics, nanowires, molecular chains, polarons, DNA, charge transfer, Holstein model.

\section*{INTRODUCTION}


Elucidation of the mechanisms of electron transport in DNA which is the most important problem of nanobioelectronics is dealt with in a large number of theoretical and experimental works \cite{Chetverikov} -- \cite{Lak2000}. Of particular interest is the study of charge transfer in DNA in the presence of an electric field \cite{PhE} -- \cite{Vinogr2019}.

In the work presented, the motion of a charge along a chain in the presence of a constant electric field is simulated on the basis of the Holstein model \cite{Holstein,Holstein2}. Modeling of the charge motion even in a homogeneous polynucleotide chain is a multipleparameter problem. Therefore, despite the simplicity of the model chosen, various and complex dynamic regimes can take place in the system under consideration. The character of the motion and distribution of a charge along the chain depends on many parameters of the system: on each of the parameters of the chain, on the value of the electric field intensity, on the initial distribution of the charge in the chain.

Earlier studies (\cite{PhE}, \cite{LaKo2011}, \cite{Korshunova2020}) show that the uniform motion of a polaron along a chain is possible for small values of the electric field intensity. In \cite{jtf2} a possibility of a uniform motion of a polaron in a homogeneous Holstein chain in a constant electric field over very large distances was shown. For large values of the electric field intensity, a uniform motion is not observed, the charge loses its original shape and moves along the chain in the direction of the field performing Bloch oscillations. But the character of a charge motion along the chain depends not only on the value of the electric field intensity, but also on the parameters of the chain and even on the initial distribution of the charge.

In this work, we investigate the motion of a charge in a constant electric field in polynucleotide chains with different values of the constant of coupling between the charge and the displacements of the chain. The values of other parameters of the chain do not change. In the course of a uniform motion along the chain, the charge moves maintaining its shape. In this case, we can say that a polaron is moving along the chain. In this work, the dependence of the velocity of a uniformly moving polaron on the value of the coupling constant between the charge and the displacements of the chain is shown. In the course of oscillatory motion, the charge loses its original shape and moves along the chain in the direction of the field, performing Bloch oscillations. In this case, the instantaneous and average velocities of the charge motion were considered and their dependence on the value of the coupling constant was investigated.

\section*{MATHEMATICAL MODEL AND INITIAL DATA}

Simulation of a charge motion in a homogeneous molecular chain in the presence of a constant electric field was carried out on the basis of the Holstein model \cite{Holstein,Holstein2}. Within the framework of this model, DNA is considered as a homogeneous chain composed of $N$ sites. Each site is a nucleotide pair, which is considered as a harmonic oscillator \cite{Lak2000}. The motion of a charge in a constant electric field is modeled by a system of coupled quantum-classical dynamic equations with dissipation. The dynamics of an electron is described by the linear Schr\"{o}dinger equation, and the dynamics of sites with allowance for dissipation is described by the classical equations of motion.

To simulate the dynamics of a quantum particle in a chain of $N$ nucleotide pairs, we will use the Holstein Hamiltonian, in which each site is a diatomic molecule:
To simulate the dynamics of a quantum particle in a chain of $N$ nucleotide pairs, we will use the Holstein Hamiltonian, in which each site is a diatomic molecule:
\begin{eqnarray}\label{1}
&\hat{H}=-\sum_{n}^{N}\upnu\Big(|n\rangle\langle n-1|+|n\rangle\langle n+1|\Big) + \sum_{n}^{N}\upalpha q_n|n\rangle\langle n| \nonumber \\
&+\sum_{n}^{N}M\dot{q}_n^2/2 + \sum_{n}^{N}kq_n^2/2 + \sum_{n}^{N}e\mathcal{E}n|n\rangle\langle n|,
\end{eqnarray}
where $\upnu$ is a matrix element of a charge transition between neighboring sites (nucleotide pairs),  $\upalpha$ is a constant of charge interaction with sites displacements $q_n$, $M$ is the effective mass of a site, $k$ is an elastic constant, $e$ is the electron charge, $\mathcal{E}$ is the electric field intensity.

The equations of motion for the Hamiltonian $\hat{H}$ lead to the following system of differential equations:
\begin{eqnarray}
i\hbar\dot{b}_n&=&-\upnu(b_{n-1}+b_{n+1})+\upalpha q_n b_n+e\mathcal{E}anb_n,\label{2}\\
M\ddot{q}_n&=&-\upgamma\dot{q}_n-kq_n-\upalpha|b_n|^2\,,\label{3}
\end{eqnarray}
where $b_n$ is the amplitude of the probability of a charge occurrence on the $n$-th site, $\sum_n |b_n|^2=1$, $\hbar=h/2\pi$, $h$ is Planck's constant. Classical motion equations \eqref{3} involve dissipation determined by the friction coefficient $\upgamma$.

Equations \eqref{2} are Schr\"{o}dinger equations for the probability amplitudes $b_n$, which describe evolution of a particle in deformable chain. Equations \eqref{3} are classical motion equations which describe the dynamics of nucleotide pairs with account taken for dissipation.

For numerical simulation of the polaron motion, we pass on to dimensionless variables using the relations:
\begin{eqnarray}\label{4}
\nonumber \upeta&=&\uptau\upnu/\hbar\,,\ \ \ \upomega^2=\uptau^2K/M\,, \ \ \  \upomega'\uptau\upgamma/M\,,\ \ q_n=\upbeta u_n, \\
E&=&\mathcal{E}ea\uptau\big/\hbar,
\ \  \upkappa\upomega^2=\uptau^3(\upalpha)^2/M\hbar\,,\;
\upbeta=\uptau^2\upalpha/M\,,\ \ \ t=\uptau\tilde{t}\,,
\end{eqnarray}
where $\uptau$ is an arbitrary time scale which relates time $t$ and dimensionless variable $\tilde{t}$, \  $\tilde{t}=t/\uptau$, \ $\uptau=10^{-14}$ sec (arbitrary time scale).

In dimensionless variables \eqref{4} equations \eqref{2}, \eqref{3} take the form:
\begin{eqnarray}
i\frac{db_n}{d\tilde{t}}&=&-\upeta\bigl(b_{n+1}+b_{n-1}\bigr)+\upkappa\upomega^2u_nb_n+Enb_n, \label{5}\\
\!\!\!\frac{d^2u_n}{d\tilde{t}^2}\!&=&\!-\upomega'\frac{du_n}{d\tilde{t}}-\upomega^2u_n-|b_n|^2\,,\label{6}
\end{eqnarray}
where $b_n$ are amplitudes of the probability of charge's occurrence on the $n$-th site, $\sum_n |b_n(\tilde{t})|^2=1$,  $\upeta$ -- are matrix elements of the transition through sites, $\upomega$ is the frequency of oscillations of the $n$-th site,  $\upkappa$ is the coupling constant, $\upomega'$ is a friction coefficient, $u_n$ are displacements of sites from their equilibrium positions,   $E$ is the electric field intensity. In dimensional units the electric field intensity is $\mathcal{E} \approx E \cdot 1.88 \cdot 10^6V/cm$.

The model introduced in this way, which describes the dynamics of a charged particle in a polynucleotide chain, explicitly takes into account dissipation in the system under consideration.

The system of nonlinear differential equations \eqref{5}, \eqref{6} is solved by the Runge-Kutta method of the 4th order. The calculations were carried out using the computing facilities of the JSCC RAS.

In the absence of an electric field, the system of equations \eqref{5}, \eqref{6} in the continual limit has a stationary solution in the form of an inverse hyperbolic cosine:
\begin{eqnarray}\label{7}
|\,b_n(0)|=\frac{\sqrt{2}}{4}\sqrt{\frac{\upkappa}{|\,\upeta|}}\,\mathrm{ch}^{-1}
\Bigl(\frac{\upkappa(n-n_0)}{4|\,\upeta|}\Bigr),\\
u_n(0)=|\,b_n(0)|^2\big/\upomega^2,\ \  du_n(0)\big/d\tilde{t}=0. \nonumber
\end{eqnarray}

In this work, we investigate the motion of a charge in a constant electric field in polynucleotide chains with different values of the coupling constant $\upkappa$. To simulate the motion of a charge, we will use the fixed values of the following parameters: matrix elements of the transition along the sites $\upeta=2.4$, the oscillation frequency of the sites $\upomega=1$, and the friction coefficient $\upomega'=1$.

To simulate the motion of a polaron in a constant electric field, we will place an initial polaron of the form \eqref{7} in the chain. We place the center of the polaron on the site with the number $n_0$. The value of $n_0$  is chosen so that at the beginning of the calculations the polaron be far enough from the ends of the chain. Similarly, the length of the chain is chosen so that at the end of the calculations the polaron would not come too close to the end of the chain. The motion of a charge in an electric field is modeled in a homogeneous open chain with two ends. The field turns on "instantly" at the initial moment of time.

\begin{figure}[h]
	\centering
	\resizebox{0.56\textwidth}{!}{\includegraphics{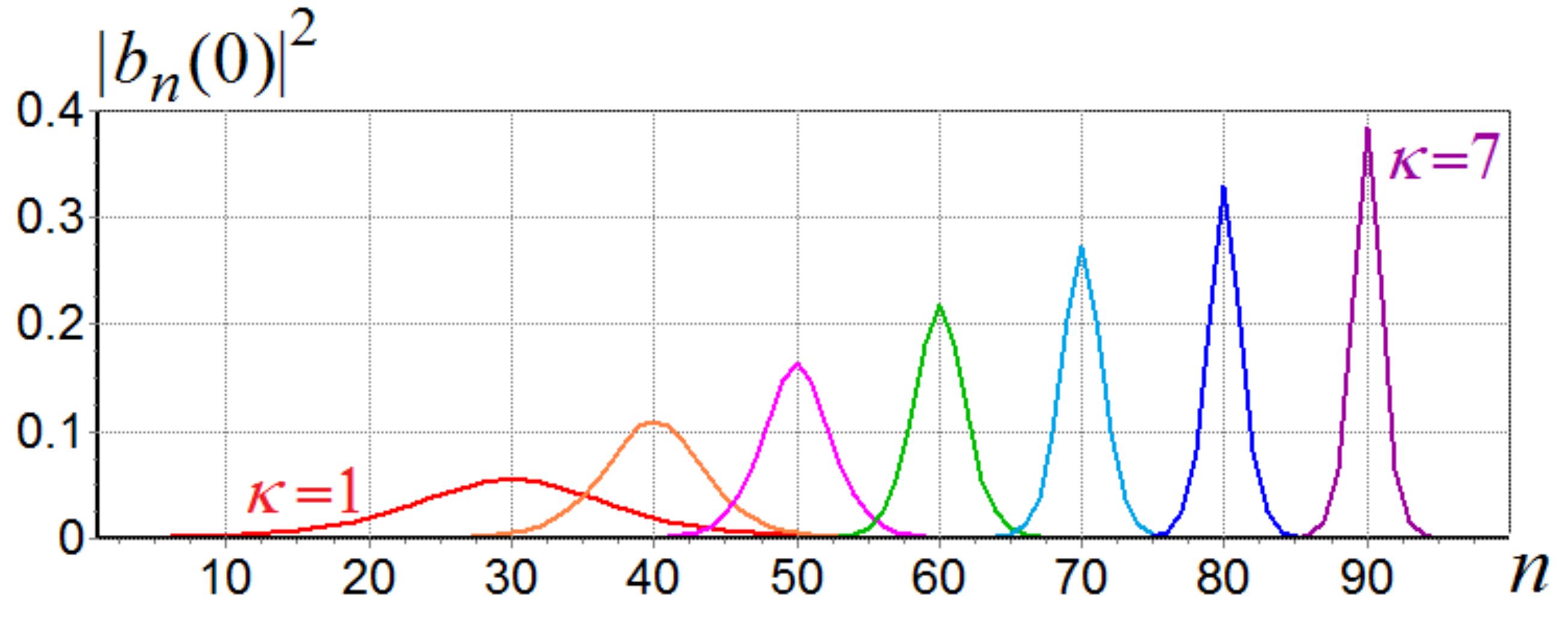}}
	\caption{Graphs of the functions $|b_n(0)|^2$ of the form \eqref{7} in a chain of $N=100$ sites for different values of the coupling constant $\upkappa=1, 2, 3, 4, 5, 6, 7.$}\label{Bn0_1-7}
\end{figure}

Figure \ref{Bn0_1-7} shows seven graphs of the functions $|b_n(0)|^2$ of the form \eqref{7} in a chain of $N=100$ sites for different values of the coupling constant $\upkappa=1, 2, ..., 7$. In the example presented, the following dimensionless values of the chain parameters were chosen: $\upeta=2.4, \upomega=1, \upomega'=1$. The graphs of functions $|b_n(0)|^2$ shown in \ref{Bn0_1-7} clearly demonstrate the dependence of the initial polaron state of the form \eqref{7} on the value of the coupling constant $\upkappa$.

Functions $|b_n(0)|^2$ of the form \eqref{7} are a solution of the system of equations \eqref{5}, \eqref{6} in the continuum limit, but for discrete chains, this solution is only an approximation to the function $|b_n|^2$ 2 in a discrete chain in the absence of an electric field and external excitations. That is, the form of a polaron in a discrete chain is slightly different from the initial polaron state of the form  \eqref{7}. For chains with small values of the coupling constant $\upkappa\leq4$ for given values of the matrix elements of the transition through the sites $\upeta=2.4$ the functions  $|b_n(0)|^2$ of the form \eqref{7} practically coincide with the shape of a polaron in a discrete chain. In chains with a coupling constant  $\upkappa\ge5$ the discrete polaron is slightly higher and narrower than the function $|b_n(0)|^2$ of the form \eqref{7}, but in this study it is not of fundamental importance, so we will use the function $|b_n(0)|^2$ of the form \eqref{7} as the initial polaron state.

In the example in Fig. \ref{Bn0_1-7}, as in the examples below, the sites in the chain are numbered from left to right. We set the values of the electric field intensity to be positive: $\widetilde{E}>0$, the charges move along the chain in the direction of the field from right to left, therefore the values of $n_0$ (the position of the initial polaron) are set near the right-hand end of the chain.

\section*{UNIFORM POLARON MOTION AND OSCILLATORY REGIME OF A CHARGE MOTION}

\begin{figure}[h]
	\centering
	\resizebox{0.52\textwidth}{!}{\includegraphics{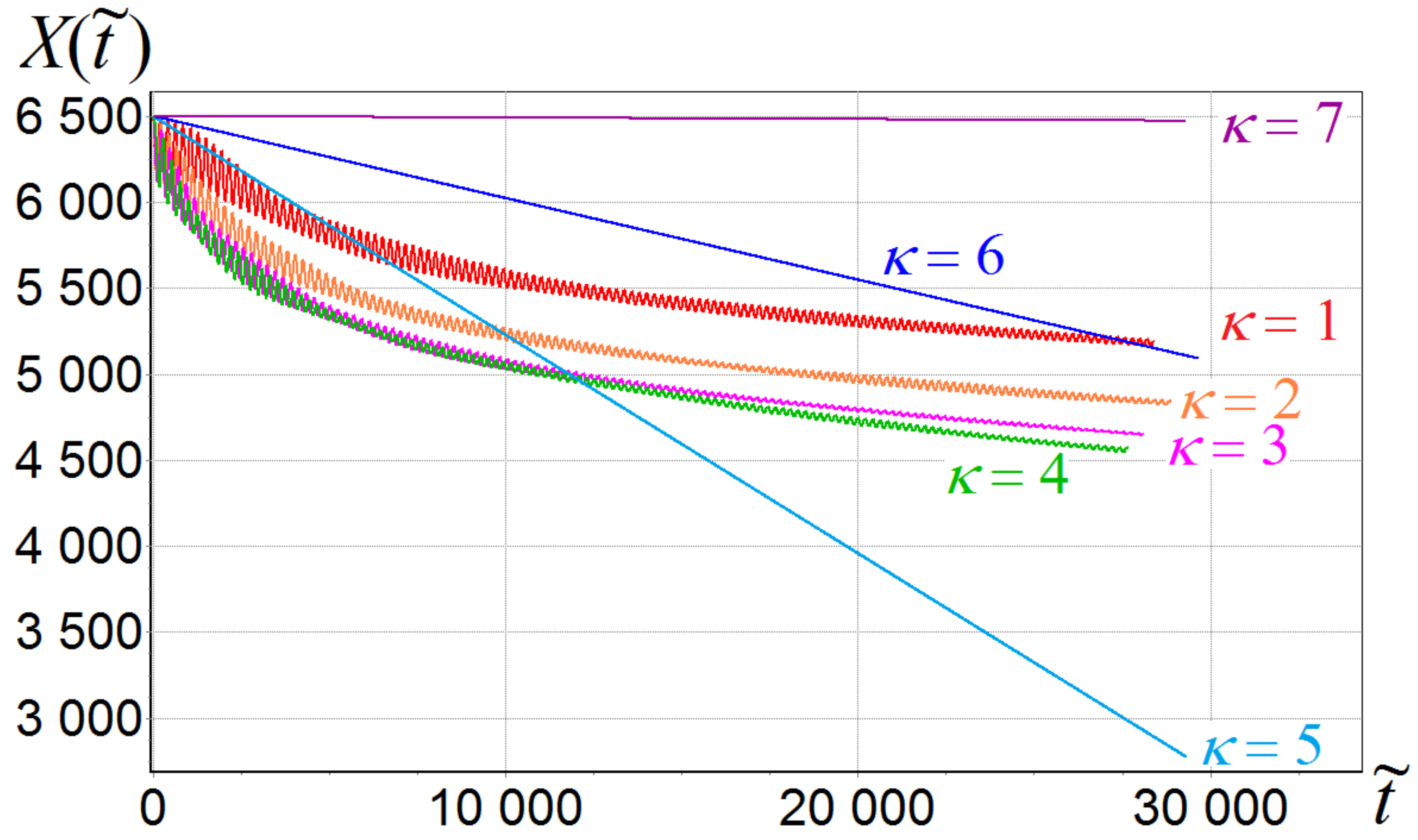}}
	\caption{Graphs of functions $X(\widetilde{t})$ for different values of the coupling constant $\upkappa=1, 2, ..., 7$. The chain length is $N=7001$ The center of the initial polaron state of the form \eqref{7} $n_0=6500$. The values of the chain parameters are: $\upeta=2.4, \upomega=1, \upomega'=1$. The value of the electric field intensity is $E=0.03$.}\label{X(t)_0.03_kp_1-7}
\end{figure}

Figure \ref{X(t)_0.03_kp_1-7} shows the examples of the evolution of a charge from the initial polaron state of the form \eqref{7} in seven chains with different values of the coupling constant $\upkappa$. The graphs of the initial polaron state of the form \eqref{7} for each value of the coupling constant $\upkappa$ are shown in Fig. \ref{Bn0_1-7}. As mentioned above, the values of the following parameters are set to be the same in all the chains: matrix elements of the transition through sites $\upeta=2.4$, oscillation frequency of sites
$\upomega=1$, friction coefficient $\upomega'=1$. The initial values of $|b_n(0)|$  are given in the form of an inverse hyperbolic cosine of the form \eqref{7}, the center of which is located at the site of the chain with the number $n_0=6500$. The length of the chain is  $N=7001$ sites.

Figure \ref{X(t)_0.03_kp_1-7} shows the graphs of the function $X(\widetilde{t})$, which describes the position of the center of mass of a charge for the value of the electric field intensity $E=0.03$, $X(\widetilde{t})~=~\sum\nolimits_{n}{|\,b_n(\widetilde{t})|^2}\cdot n$.

The graphs of the function $X(\widetilde{t})$ in Fig. \ref{X(t)_0.03_kp_1-7} in chains with large values of $\upkappa=5, 6, 7$   demonstrate a linear dependence on $\widetilde{t}$ for the specified value of the electric field intensity $E=0.03$ during the dimensionless time $\widetilde{t}\approx 30000$, therefore, for the chosen value of the electric field intensity, a uniform motion is observed in chains with the specified values of the coupling constant $\upkappa$, while the charge moves along the chain, retaining its original shape, that is, in these cases, a uniform motion of the polaron is observed. The graphs of the functions $X(\widetilde{t})$ presented in Fig.\ref{X(t)_0.03_kp_1-7} for $\upkappa=5, 6, 7$ suggest that the velocity of polaron motion in the course of its uniform motion along the chain increases as the coupling constant $\upkappa$ decreases.

In chains with smaller values of the coupling constant $\upkappa=1, 2, 3, 4$, the charge immediately starts an oscillatory motion. In what follows we will show that the charge performs Bloch oscillations. In the course of oscillatory motion along the chain, the charge quickly loses its original shape and, being distributed along the chain, moves in the direction of the field. The graphs of the functions$X(\widetilde{t})$ for $\upkappa=1, 2, 3, 4$ in Fig. \ref{X(t)_0.03_kp_1-7} show that it would be incorrect to compare the velocity of an oscillatorily moving charge with that of a uniformly moving polaron for $\upkappa=5, 6, 7$. We can only notice that during the computation time shown in Fig. \ref{X(t)_0.03_kp_1-7}, the charge with a large value of $\upkappa=4$ has travelled a greater distance along the chain.  As the coupling constant  $\upkappa$ decreases, for $\upkappa\to 0$, the total velocity of the charge tends to zero, the charge, maintaining its shape, performs Bloch oscillations near the center of the initial position of the charge, being located at the chain sites, whose number is approximately equal to one maximum Bloch amplitude for a given value of the electric field intensity.  This corresponds to the fact that in a rigid chain, for $\upkappa=0$, the charge performs Bloch oscillations maintaining its form/shape and position in the chain, that is, at the end of each oscillation period, the charge returns to its initial position. In a deformed chain, when the charge interacts with the displacements of the chain, for $\upkappa>0$, the charge loses its original shape and, performing Bloch oscillations, moves along the chain in the direction of the field, see \cite{Fi_Lak_2004}.

Thus, the total velocity of a charge during its oscillatory motion decreases with a decrease in the coupling constant $\upkappa$, while the velocity of a uniformly moving polaron increases with a decrease in $\upkappa$.

\begin{figure}[h]
	\centering
	\resizebox{0.54\textwidth}{!}{\includegraphics{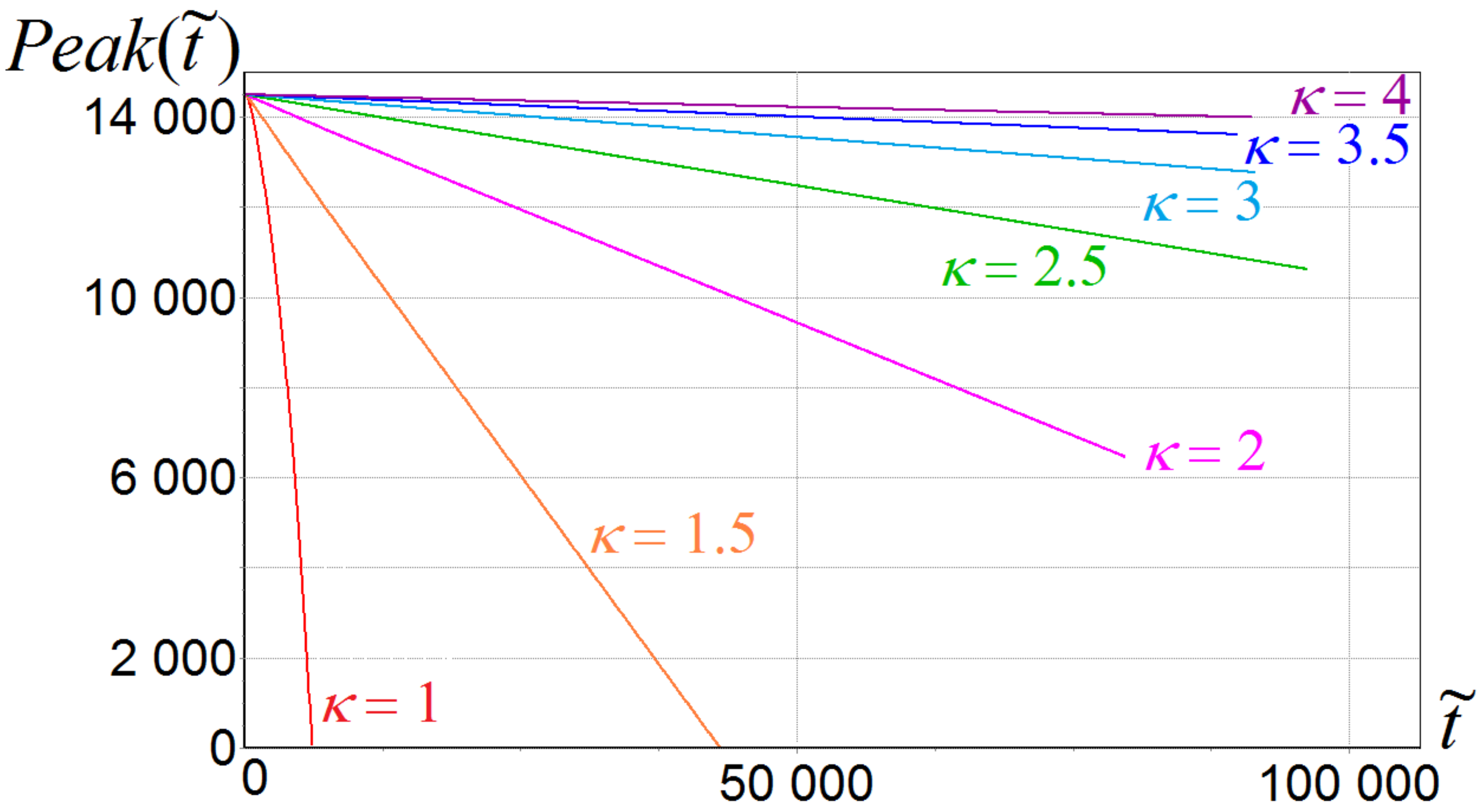}}
	\caption{Graphs of the functions $Peak(\widetilde{t})$ for different values of the coupling constant $\upkappa=1, 1.5, 2, 2.5, 3, 3.5, 4$. The chain length is $N=15001$ sites. The center of the initial polaron state of the form \eqref{7} is $n_0=14500$. The values of the chain parameters are: $\upeta=2.4, \upomega=1, \upomega'=1$. The value of the electric field intensity is $E=0.0005$.}\label{PIK_E_0_0005}
\end{figure}

Figure \ref{PIK_E_0_0005} demonstrates some examples of a uniform motion of a polaron in chains with different values of the coupling constant: $\upkappa=1, 1.5, 2, 2.5, 3, 3.5, 4$. The electric field intensity is $E=0.0005$. The center of the initial polaron state of the form \eqref{7} is located at the site $n_0=14500$ in a chain consisting of $N=15001$ sites. The initial polaron of the form \eqref{7} moves uniformly along the chain keeping its shape. Therefore, Fig. \ref{PIK_E_0_0005} shows the graphs of the functions  $Peak(\widetilde{t})$, which demonstrate the position of the polaron peak, or the number of the site at which the maximum of the function $|b_n(\widetilde{t})|^2$ occurs. With this mode of motion, the graphs of the functions  $Peak(\widetilde{t})$ practically coincide with the graphs of the functions  $X(\widetilde{t})$.

The graphs of the functions $Peak(\widetilde{t})$ shown in \ref{PIK_E_0_0005} clearly demonstrate that the velocity of a uniform motion of a polaron along the chain increases sharply as the coupling constants $\upkappa$. decreases.
Note also that in chains with large values of the coupling constant $\upkappa>4$, the initial polaron does not shift from its initial position at the electric field intensity $E=0.0005$. Fig. \ref{X(t)_0.03_kp_1-7} shows that when the electric field intensity is  $E=0.03$ the initial polaron does not shift in a chain with $\upkappa\ge7$.

\begin{figure}[h]
	\centering
	\resizebox{0.53\textwidth}{!}{\includegraphics{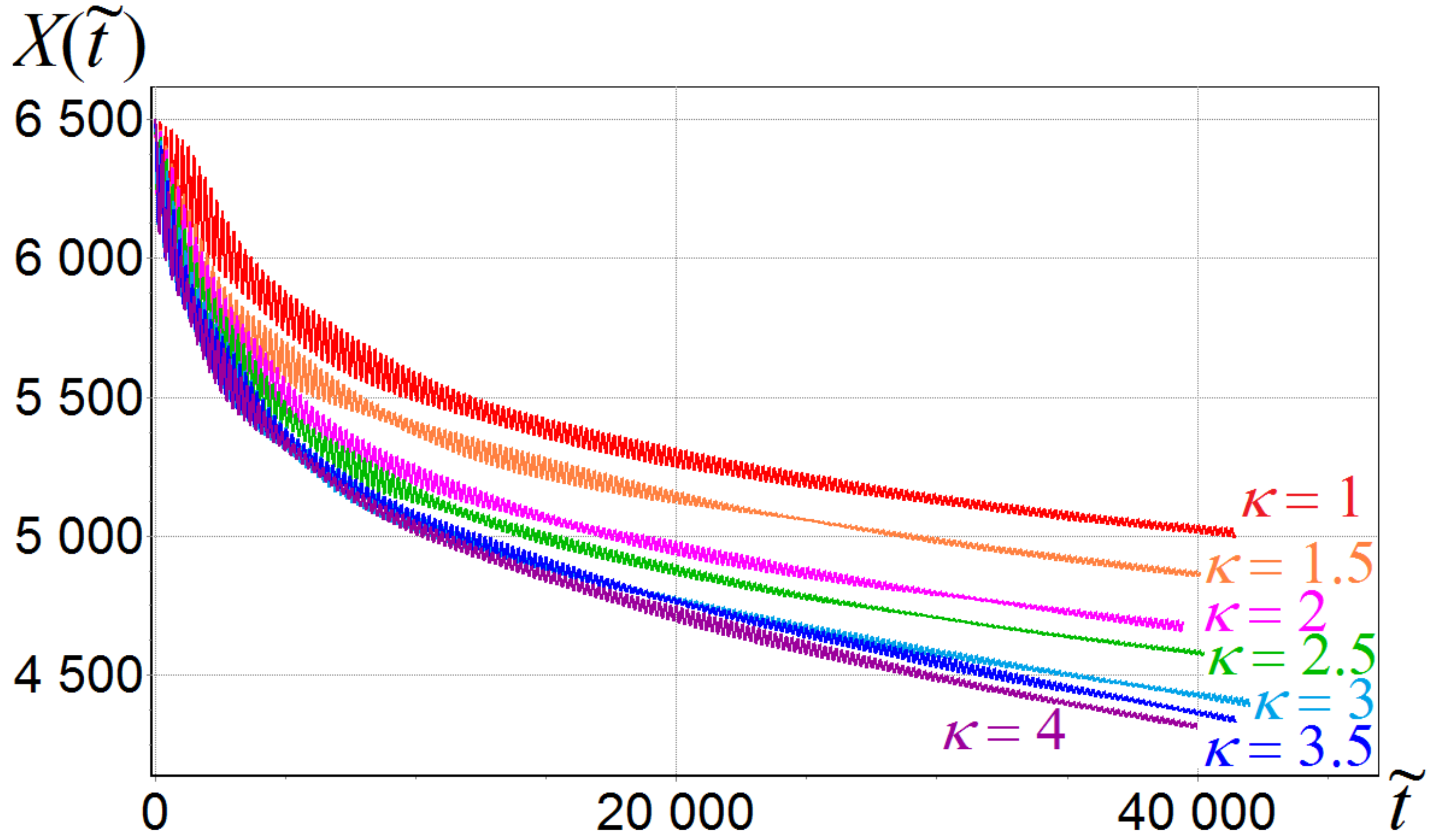}} \textbf{a)} \\
	\resizebox{0.4\textwidth}{!}{\includegraphics{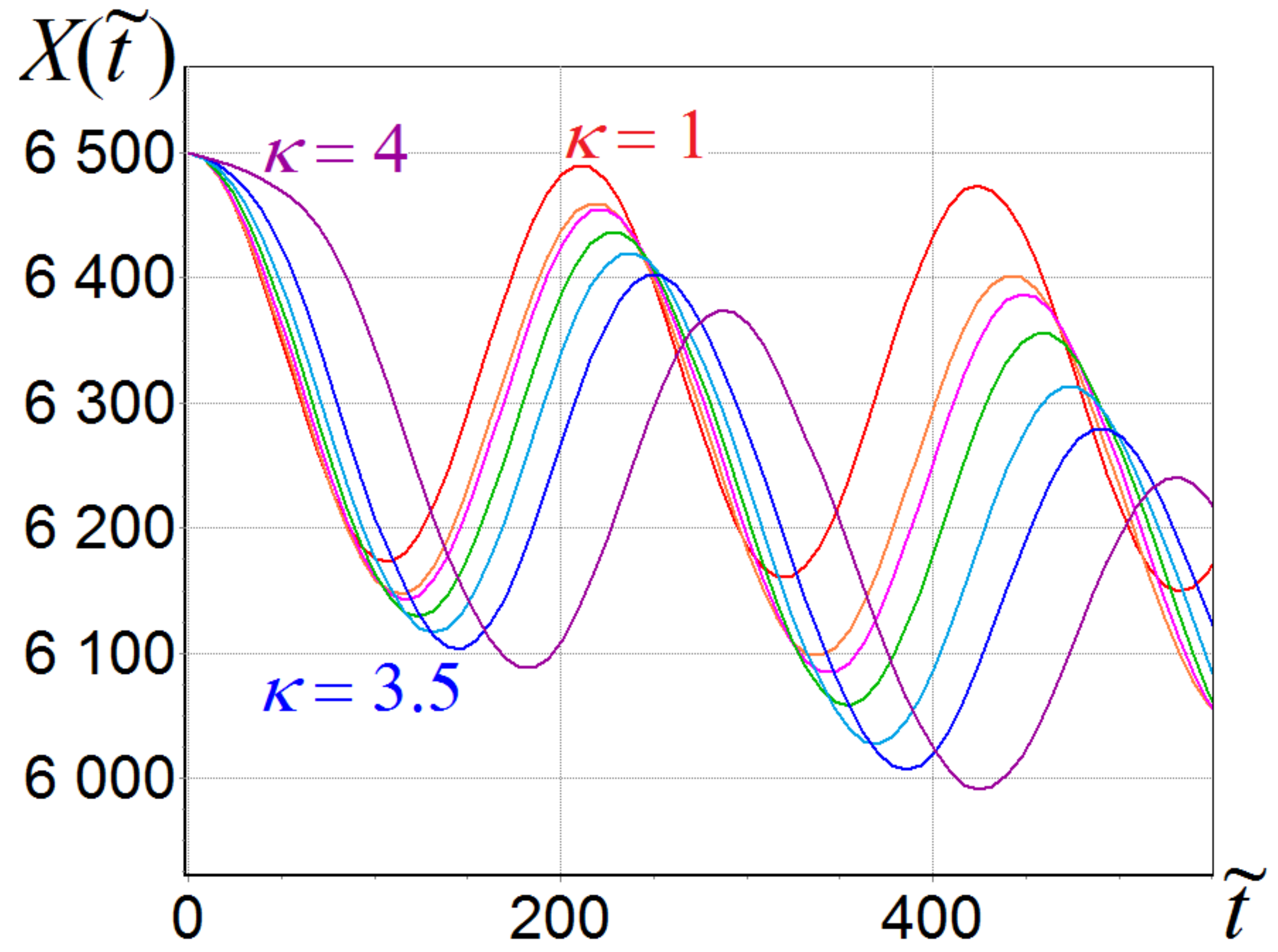}} \textbf{b)}\hspace{0.5cm}
	\resizebox{0.4\textwidth}{!}{\includegraphics{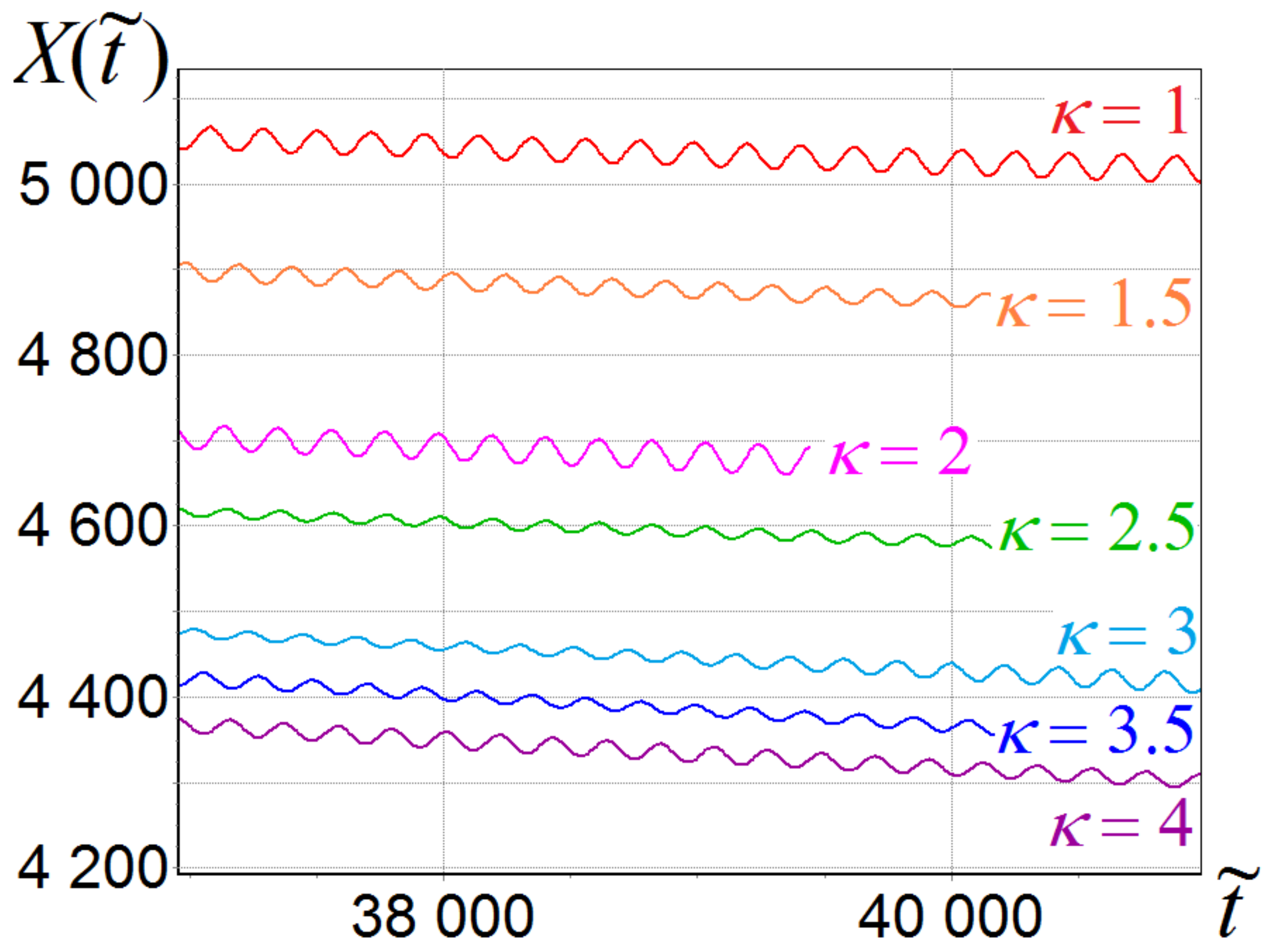}} \textbf{c)}
	\caption{Graphs of the functions $X(\widetilde{t})$ for different values of the coupling constant $\upkappa=1, 1.5, 2, 2.5, 3, 3.5, 4$. The chain length is $N=7001$ sites. The center of the initial polaron state of the form \eqref{7} is $n_0=6500$. The values of the chain parameters are $\upeta=2.4, \upomega=1, \upomega'=1$. The value of the electric field intensity is $E=0.03$. In fig. b) the graphs of the functions $X(\widetilde{t})$ are shown at the initial period of time; in fig. c) -- at the end of calculations.  }\label{X(t)_0.03}
\end{figure}

\begin{figure}[h]
	\centering
	\resizebox{0.45\textwidth}{!}{\includegraphics{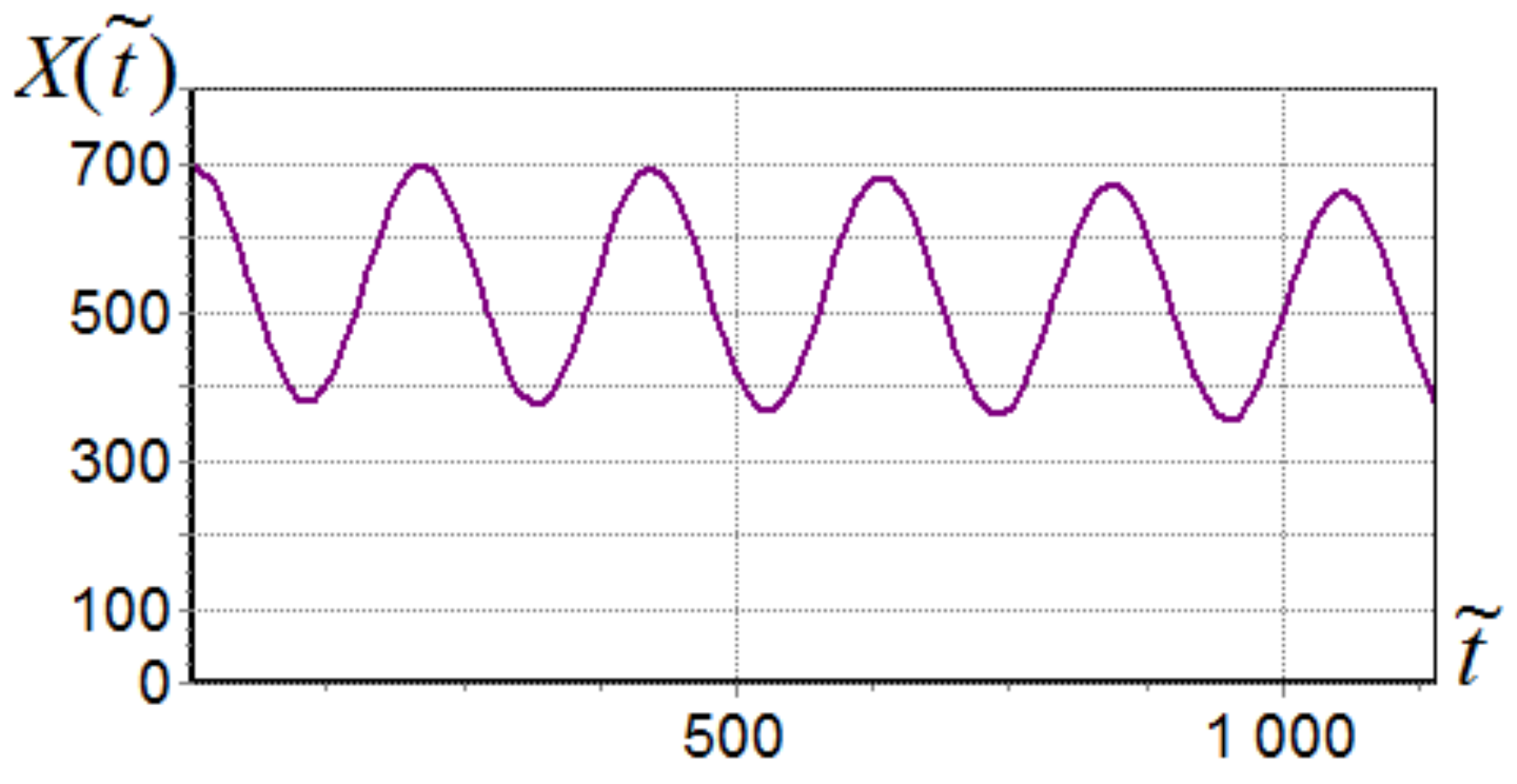}} \textbf{\small{a)}}
	\resizebox{0.45\textwidth}{!}{\includegraphics{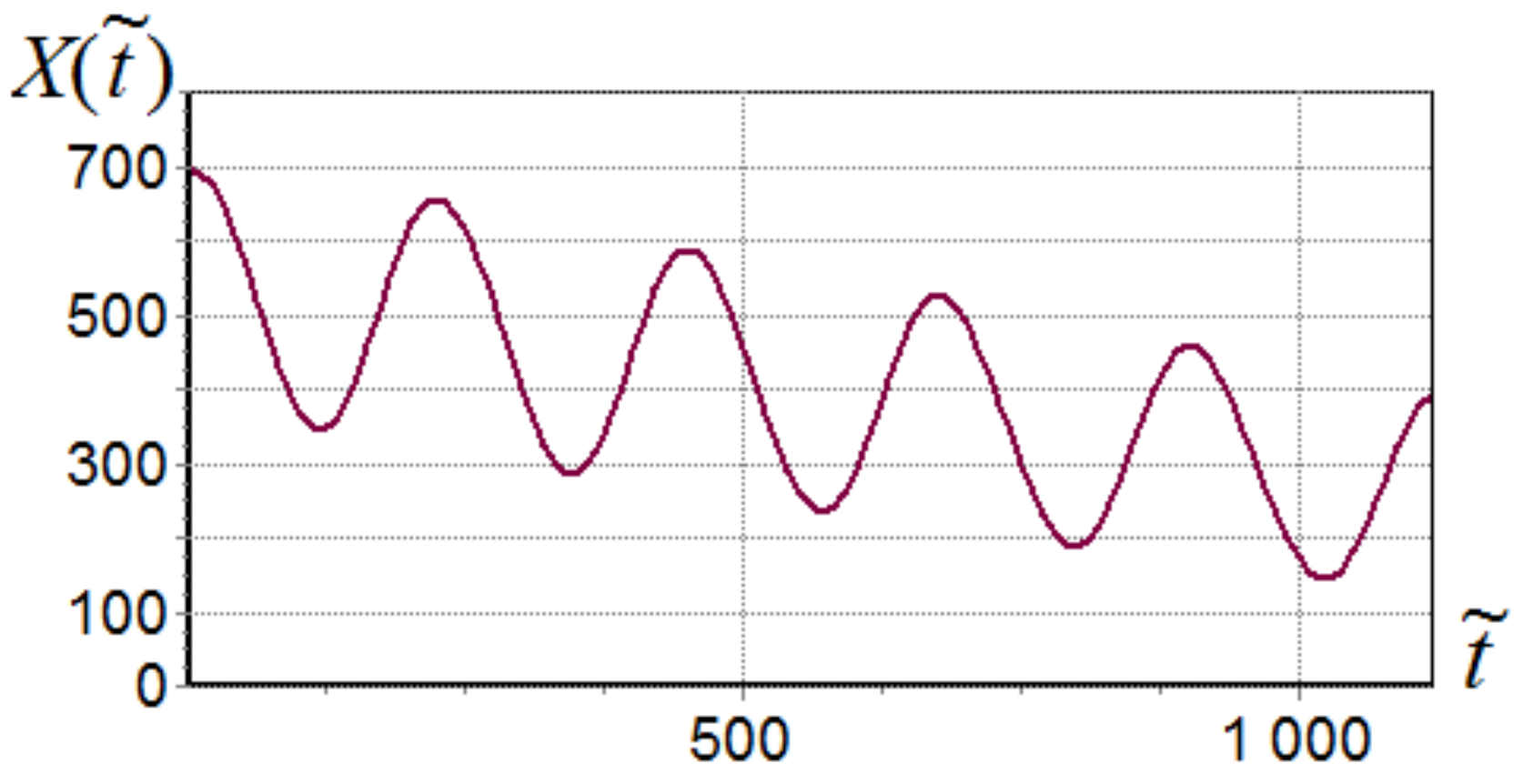}} \textbf{\small{c)}}
	\resizebox{0.45\textwidth}{!}{\includegraphics{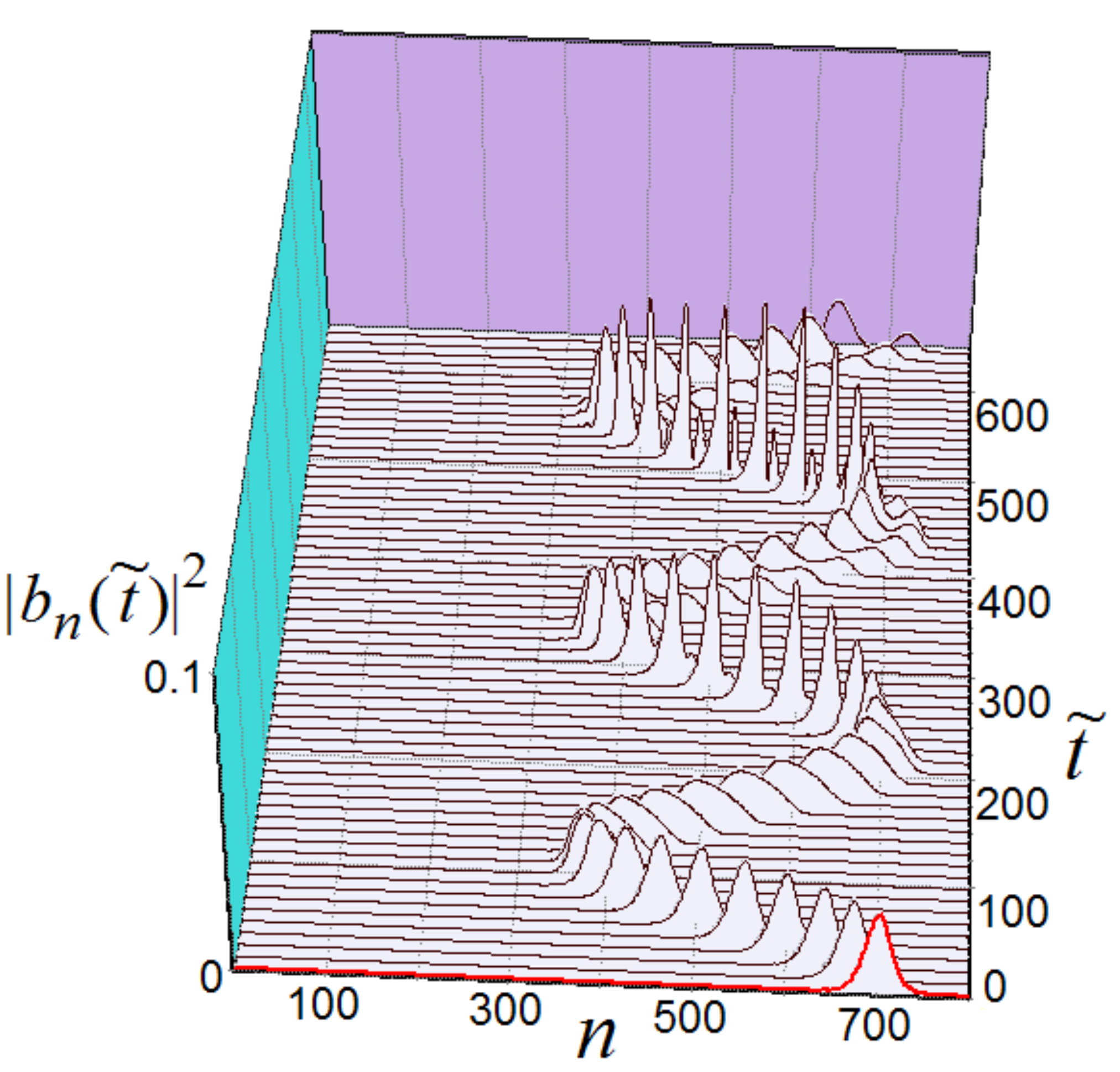}} \textbf{\small{b)}}
	\resizebox{0.45\textwidth}{!}{\includegraphics{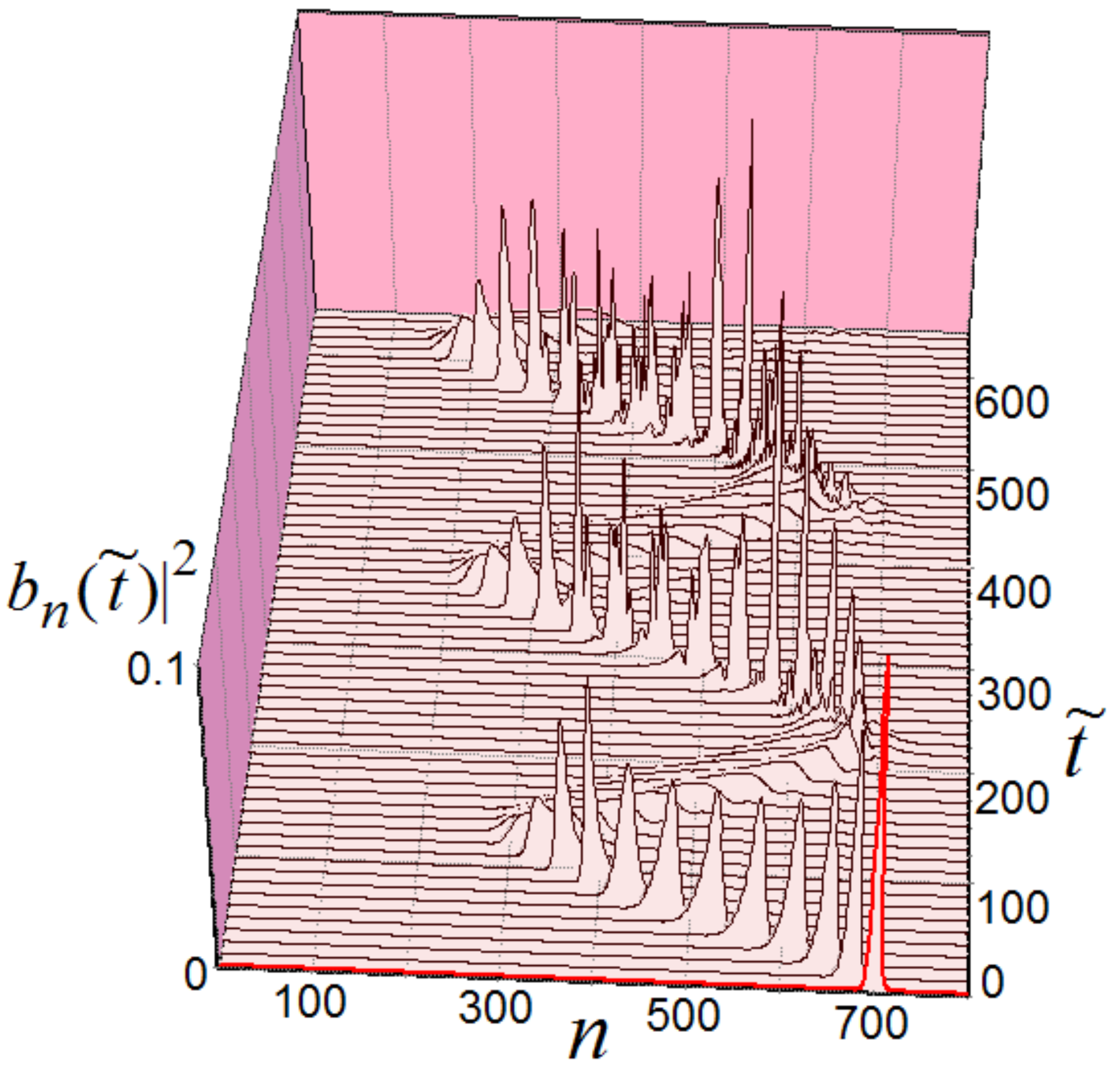}} \textbf{\small{d)}}
 	\caption{Graphs of the functions $X(\widetilde{t})$ and $|b_n(\widetilde{t})|^2$ in the course of the oscillatory motion of a charge in the chains with different values of the coupling constant. In figures a) and b) $\upkappa=0.5$, in figures c) and d) $\upkappa=2$. The chain length is $N=801$ sites. The center of the initial polaron state of the form \eqref{7} is $n_0=700$. The values of the chain parameters are: $\upeta=2.4, \upomega=1, \upomega'=1$. The value of the electric field intensity is $E=0.03$.}\label{X(t)_Bn_0.03}
\end{figure}

Figures \ref{X(t)_0.03} and \ref{X(t)_Bn_0.03} illustrate the oscillatory regime of the charge motion in a constant electric field of intensity $E=0.03$. For the intensity value $E=0.03$ the period of Bloch oscillations is $T_{BL}=2\pi/E\approx209$. The maximum Bloch amplitude $A_{BL}=4\upeta/E \approx320$. The maximum charge velocity in the course of Bloch oscillations $V_{BL}=2\upeta\approx4.8$.
The graphs of the functions $X(\widetilde{t})$ and $|b_n(\widetilde{t})|^2$ shown in Fig. \ref{X(t)_0.03} and Fig. \ref{X(t)_Bn_0.03} demonstrate good agreement between the numerical and theoretical characteristics of Bloch oscillations.

Fig. \ref{X(t)_0.03} shows some examples of the oscillatory regime of a charge motion in chains with the same values of the coupling constant as in Fig. \ref{PIK_E_0_0005}: $\upkappa=1, 1.5, 2, 2.5, 3, 3.5, 4$.
In this case, the charge loses its original shape/form and, being distributed along the chain, moves in the direction of the field. When the value of the electric field intensity is $E=0.03$ no uniform motion in chains with the indicated values of the coupling constant is observed. For this value of the electric field intensity $E=0.03$ the charge can move uniformly in chains with a coupling constant $\upkappa=5$ and larger, see Fig. \ref{X(t)_0.03_kp_1-7}. The graphs of the functions  $X(\widetilde{t})$ in Fig. \ref{X(t)_0.03} a) and b) show that the charge quickly passes into an oscillatory regime of motion. In a chain with the largest value of the coupling constant $\upkappa=4$ the charge starts an oscillatory motion with a certain delay. The charge loses its shape not instantaneously. This leads to an increase in the oscillation period and the maximum amplitude of the charge displacement along the sites in the initial period of time, see fig. \ref{X(t)_0.03}b). But as time goes on, the period of charge oscillations in all the chains with the values of the coupling constant $\upkappa$ chosen in Fig. \ref{X(t)_0.03} becomes close to the theoretical value of the period of Bloch oscillations for a given value of the electric field intensity $E=0.03$ -- $T_{BL}\approx209$.

The theoretical values of the main characteristics of Bloch oscillations do not depend on the value of the coupling constant $\upkappa$. In discrete chains, the maximum amplitude and the period of oscillations differ slightly from the corresponding theoretical values when the coupling constant $\upkappa$ changes. This fact is clearly seen in Fig. \ref{X(t)_0.03}b): in a chain with a coupling constant  $\upkappa=1$ the oscillation period and the maximum oscillation amplitude almost exactly coincide with the relevant characteristics of Bloch oscillations for a given value of the electric field intensity. Thus, it can be seen that the smaller is the value of the coupling constant in the chain, the closer are the characteristics of charge oscillations to the corresponding characteristics of Bloch oscillations.

The examples in Fig.  \ref{X(t)_0.03} represent charge oscillations in different chains. Therefore, we can consider the instantaneous charge velocity and the average velocity. The maximum instantaneous charge velocity in all the chains in Fig. \ref{X(t)_0.03}b) almost exactly coincides with the theoretical value of the maximum charge velocity in the course of Bloch oscillations  $V_{BL}=2\upeta\approx4.8$.

In Figure \ref{X(t)_0.03}c), the graphs of the functions $X(\widetilde{t})$ are shown at the end of the computation period. Obviously, the value of $X(\widetilde{t})$ is not equal to the site reached by the charge in the chain, since the function  $X(\widetilde{t})$ describes the position of the center of mass of the charge. In all the examples shown in Fig. \ref{X(t)_0.03}c) the charges did not reach the end of the chains, since all the graphs of the functions $X(\widetilde{t})$ demonstrate the same oscillation period, close to the Bloch oscillation period. If the charge had reached the edge of the chain, then the oscillations would quickly fail, the period of oscillations would immediately begin to decrease and completely disappear. Thus, the examples in Fig. \ref{X(t)_0.03} show that the charge moves along the chain in the direction of the field, performing Bloch oscillations. In chains with greater values of the coupling constant, the charge travels a greater distance. The average charge velocity decreases in the course of motion, at the end of the calculation period shown (see Fig.  \ref{X(t)_0.03}c)) the average charge velocity in all the chains is almost the same and, obviously, it will further approach zero.

Figure \ref{X(t)_Bn_0.03} shows two examples of the oscillatory regime of the charge motion in the initial period of time in chains with different values of the coupling constant. Figures \ref{X(t)_Bn_0.03}a) and \ref{X(t)_Bn_0.03}b) demonstrate a chain with a coupling constant $\upkappa=0.5$. Figures \ref{X(t)_Bn_0.03}c) and \ref{X(t)_Bn_0.03}d) -- a chain with a coupling constant  $\upkappa=2$. The value of the electric field intensity is the same as in the previous example: $E=0.03$. The center of the initial polaron state of the form  \eqref{7} is located at the site $n_0=700$ in a chain consisting of  $N=801$ sites.

In a chain with a coupling constant $\upkappa=0.5$, see Fig. \ref{X(t)_Bn_0.03}b), the initial polaron state of the form \eqref{7} is much wider than the initial polaron state in the chain with the coupling constant $\upkappa=2$, see Fig. \ref{X(t)_Bn_0.03}d). In a chain with a coupling constant $\upkappa=2$ the charge quickly loses its original shape, the period of charge oscillations and the maximum oscillation amplitude being sufficiently close to the corresponding characteristics of Bloch oscillations.

In a chain with a small value of the coupling constant $\upkappa=0.5$, Fig. \ref{X(t)_Bn_0.03}a) and b), the broad initial polaron state moves along the chain keeping its shape during the first oscillation period, then the charge gradually loses its original shape and, being distributed along the chain, moves on average in the direction of the field, performing Bloch oscillations. In this case, the period of charge oscillations and the maximum oscillation amplitude practically coincide with the theoretical values of the period of Bloch oscillations and the maximum Bloch amplitude for a given value of the electric field intensity $E=0.03$: $T_{BL}=2\pi/E\approx209$, $A_{BL}=4\upeta/E \approx320$. As $\upkappa\to 0$ the charge, retaining its original shape, performs Bloch oscillations near the initial position of the charge. At that, the graphs of the functions  $X(\widetilde{t})$ and $|b_n(\widetilde{t})|^2$, during a large number of oscillations, are similar to the corresponding graphs during the first Bloch period shown in Fig. \ref{X(t)_Bn_0.03}a) and \ref{X(t)_Bn_0.03}b), that is, after each period of oscillations, the charge returns to its initial position.

The maximum instantaneous charge velocity in the examples in Fig. \ref{X(t)_Bn_0.03} is practically the same and is equal to the maximum charge velocity in the process of Bloch oscillations $V_{BL}=2\upeta\approx4.8$. The average charge velocity at the initial period of time is much higher in the chain with a larger value of the coupling constant $\upkappa=2$, see Fig. \ref{X(t)_Bn_0.03}a) and Fig.  \ref{X(t)_Bn_0.03}c). Recall that when a charge moves uniformly, the velocity is higher in chains with smaller values of the coupling constant $\upkappa$.

\section*{CONCLUSION}

In this work, we studied various regimes of a charge motion in a constant electric field in polynucleotide chains with different values of the coupling constant $\upkappa$. The values of the following chain parameters: matrix elements of the transitions over sites  $\upeta=2.4$, site oscillation frequency $\upomega=1$, friction coefficient $\upomega'=1$, did not changed in all the above examples.

In the course of a uniform motion, the charge moves along the chain at a constant velocity maintaining its shape, therefore, a polaron moves along the chain. The calculations showed that the velocity of a uniform motion of a polaron along the chain increases sharply with a decrease in the coupling constant $\upkappa$, that is, "wider" (see fig. \ref{Bn0_1-7}) polarons move faster.

It is shown that for a fixed value of the electric field intensity $E$,a sufficiently narrow polaron does not shift from its initial position, or, in other words, for a given value of the electric field intensity $E$, one can choose a chain with such a (sufficiently large) value of the coupling constant $\upkappa$, in which the initial polaron will remain stationary (see Fig. \ref{X(t)_0.03_kp_1-7} and Fig. \ref{PIK_E_0_0005}).

Since the character of the motion and distribution of a charge along the chain in the course of the oscillatory motion is completely different from the uniform motion of a polaron, the dependence of the character of the motion of the oscillating charge on the coupling constant $\upkappa$ is completely different. In the oscillatory mode, the charge loses its original shape and, being distributed along the chain, moves in the direction of the field, performing Bloch oscillations. It is shown that the total velocity of a charge during its oscillatory motion decreases as the coupling constant  $\upkappa$ diminishes; the charge travels a greater distance in a chain with a larger value of the coupling constant $\upkappa$. It is also shown that the maximum instantaneous charge velocity in chains with different values of the coupling constant $\upkappa$ is practically the same and is equal to the maximum charge velocity in the course of Bloch oscillations $V_{BL}=2\upeta$.

The chosen values of the parameters of the chains: $\upeta=2.4$, $\upomega=1$, $\upomega'=1$, are model. With such parameters, numerical simulation can be carried out much faster. The $Poly A / Poly T$ chain corresponds to the following dimensionless values of the chain parameters: $\upkappa=4$, $\upeta=2.4$, $\upomega=0.01$, $\upomega'=0.006$. With these parameters, much more time is required to carry out calculations. But the dependence of the character of the charge motion on the value of the coupling constant $\upkappa$ in chains with DNA parameters is similar to the corresponding dependence in the model examples considered.
In the examples in Figures  \ref{X(t)_0.03_kp_1-7}, \ref{X(t)_0.03} and \ref{X(t)_Bn_0.03} the motion of a charge in a field with a dimensionless intensity $E=0.03$ was considered. In dimensional units, this value of the electric field intensity is approximately equal to the value $\mathcal{E} \approx E \cdot 1.88 \cdot 10^6V/cm \approx 5.64 \cdot 10^4V/cm$. Preliminary calculations show that for small values of the parameters $\upomega=0.01$, $\upomega'=0.006$, which correspond to the parameters of the DNA chain, in order to carry out similar studies, it is necessary to set significantly lower values of the electric field intensity. In dimensional units, the value of the electric field intensity $\mathcal{E}$ should be of the order of $\mathcal{E} \approx 5. \cdot 10^3V/cm$ and less. As the value of the electric field intensity decreases, the velocity of a uniform motion of the polaron decreases, the amplitude and period of Bloch oscillations increase, and, as a consequence, the computational costs increase significantly.

In the future, we are planning to conduct similar studies for chains with DNA parameters and evaluate in physical terms the charge velocity and electric field intensity in various motion regimes.

{\footnotesize The work was done using the computing resources of The Joint Supercomputer Center of the Russian Academy of Sciences (JSCC RAS).}

{\footnotesize This work was supported by the RFBR grant 19-07-00406.}

\end{document}